\documentclass[letterpaper,10pt]{article}

\pdfoutput=1

\usepackage{jheppub}
\usepackage{array}
\allowdisplaybreaks

\def\ket#1{\langle #1 \rangle}

\DeclareMathOperator{\Conf}{Conf}
\DeclareMathOperator{\Gr}{Gr}
\DeclareMathOperator{\Li}{Li}

\title{Cluster Functions and Scattering Amplitudes\\ for Six and Seven Points}

\author{Thomas Harrington}
\author{and Marcus Spradlin}
\affiliation{Department of Physics, Brown University, Providence RI 02912, USA}

\abstract{Scattering amplitudes in planar super-Yang-Mills theory
satisfy several basic physical and mathematical constraints, including
physical constraints on their branch cut structure and various empirically
discovered connections to the mathematics of cluster algebras.
The power of the bootstrap program for amplitudes is inversely proportional
to the size of the intersection between these physical and mathematical
constraints: ideally we would like a list of constraints
which determine scattering amplitudes uniquely.
We explore this intersection quantitatively
for two-loop six- and seven-point amplitudes
by providing a complete taxonomy of the
$\Gr(4,6)$ and
$\Gr(4,7)$ cluster polylogarithm
functions of arXiv:1401.6446 at weight~4.
}

\begin{document}
\maketitle
\flushbottom

\section{Introduction}

Several recent papers following~\cite{Golden:2013xva} have explored
the connection between (multi-loop) scattering amplitudes
in planar $\mathcal{N}=4$ super-Yang-Mills (SYM) theory and
cluster algebras, a subject of great interest to mathematicians.
This line of research has two closely related branches: (1) investigating
purely mathematical questions having to do with
the classification of functions with certain cluster algebraic
properties, i.e.~``how rare are special functions of the type we see in SYM
theory?'', and (2)
exploiting these mathematical properties, together with physical
input as needed, to carry out calculations of new, previously intractable
amplitudes, i.e.~``how far can we get by exploiting the special properties
of cluster algebras?''.

The most basic aspect of the observed connection, supported by all
evidence available to date, is that $n$-point scattering amplitudes
in SYM theory have singularities
only at points in $\Conf_n(\mathbb{P}^3)$
(the space of massless $n$-point kinematics modulo dual conformal
invariance) where some cluster coordinate of the associated
$\Gr(4,n)$ cluster algebra vanishes.
More specifically, all known multi-loop amplitudes may be expressed
as linear combinations of generalized polylogarithm functions
written in the symbol alphabet consisting of such cluster coordinates.
We expect this to be true to all loop order for all MHV and NMHV amplitudes.

Deeper connections to the underlying cluster algebra have been found
for the two-loop MHV remainder functions $R^{(2)}_n$.  The algebra of
generalized polylogarithm functions modulo products admits
a cobracket $\delta$ satisfying $\delta^2 = 0$, giving it the
structure of a Lie coalgebra~\cite{G02}.
It has been observed that $\delta R^{(2)}_n$ has
a very rigid connection to the Poisson structure
on the kinematic domain $\Conf_n(\mathbb{P}^3)$.  Specifically,
the $(2,2)$ component of $\delta R^{(2)}_n$ can always be written
as a linear combination of $\Li_2(-x_i) \wedge \Li_2(-x_j)$ for
pairs of cluster coordinates having Poisson bracket
$\{ \log x_i, \log x_j \} = 0$, while the $(3,1)$ component can always
be written as a linear combination of $\Li_3(-x_i) \wedge \log(x_j)$ for
pairs having $\{ \log x_i, \log x_j \} = \pm 1$.
These mathematical properties are tightly constraining:  it has been
argued in~\cite{Golden:2014pua} that, when combined with a few physical
constraints, they uniquely determine the $(2,2)$ component of
$\delta R_n^{(2)}$ for all $n$.

It is an interesting open problem to determine whether (and, if so,
precisely how) the structure of more general amplitudes may be dictated
by the underlying Poisson structure on $\Conf_n(\mathbb{P}^3)$.
This is a difficult question to address because data on
multi-loop amplitudes is very hard to come by---beyond the two-loop MHV
amplitudes, explicit results for complete amplitudes at fixed loop order
are available
only for $n=6$~\cite{Dixon:2011nj,Dixon:2011pw,Dixon:2013eka,Dixon:2014voa,Dixon:2014xca,Dixon:2014iba,Dixon:2015iva}
(in addition,
the symbol of the two-loop $n=7$ NMHV amplitude has been
computed in~\cite{CaronHuot:2011kk}, and that of
the three-loop $n=7$ MHV amplitude
in~\cite{Drummond:2014ffa}).
With only a handful of results available it may be difficult to identify
a pattern which might let one tease out the underlying structure.  Moreover,
accidental simplifications may occur at small $n$ which can obscure
the general structure.  (For example, the $(2,2)$ component of
$\delta R^{(2)}_6$ is identically zero~\cite{Goncharov:2010jf}.)
It is known that the $(3,3)$ component of
$\delta R^{(3)}_6$ is not expressible in terms of cluster
$\mathcal{X}$-coordinates~\cite{Vergu}, but there could be some more deeply
hidden structure in this amplitude.

The primary goal of this paper is to
further explore the taxonomy of two-loop cluster functions, as
defined in~\cite{Golden:2014xqa}, for $n=6,7$.
We are particularly
interested in the interplay between various mathematically natural
but physically obscure conditions that certain functions can
satisfy (such as the tight cluster constraints satisfied by
all two-loop MHV amplitudes, mentioned above) and physically natural
constraints, such as the requirement that amplitudes can only
have physical branch points on the principal sheet
(the so-called ``first-entry condition''~\cite{Gaiotto:2011dt}).
In previous work including~\cite{Golden:2014pua}
it has been remarked that the mathematical and
physical constraints on MHV amplitudes seem almost orthogonal.
One of our goals here is to explore this question quantitatively by
fully classifying the dimensions of function spaces satisfying
various properties.

We begin in Section~2 with a lightning review to set some notation
and terminology.  In Sections~3 and~4 respectively we exhaustively
analyze the spaces of cluster functions on the $\Gr(4,6)$ and
$\Gr(4,7)$ cluster algebras respectively of relevance to
$n=6,7$-point amplitudes in planar SYM theory.

\section{Review and Notation}

A kinematic configuration of $n$ massless on-shell particles, with a
cyclic order (which comes naturally in gauge theories when one
looks at planar scattering amplitudes), can be parameterized in terms
of $n$ momentum twistors~\cite{Hodges:2009hk},
$Z_i \in \mathbb{P}^3$, $i=1,\ldots,n$.
The dual conformal
symmetry of planar $n$-point amplitudes in SYM theory
further implies that that they
are functions not on $(\mathbb{P}^3)^n$ but on the
smaller space $\Conf_n(\mathbb{P}^3) \cong
\Gr(4,n)/(\mathbb{C}^*)^{n-1}$~\cite{Golden:2013xva}.

Viewing each $Z_i$ as a four-component vector of
homogeneous coordinates, the
Pl\"ucker coordinates are defined
by $\langle ijkl \rangle \equiv \det(Z_i Z_j Z_k Z_l)$.
Functions on $\Conf_n(\mathbb{P}^3)$ may be written in terms of
ratios of Pl\"ucker coordinates
such as
\begin{equation}
\label{eq:example}
\frac{\langle ijkl \rangle \langle abcd \rangle}{\langle ijcd \rangle
\langle abkl \rangle}\,,
\end{equation}
or more generally in terms of ratios of homogeneous polynomials in
Pl\"ucker coordinates having total weight zero under rescaling any
of the $Z_i$.

Such objects form the building blocks for the $\Gr(4,n)$ Grassmannian
cluster algebra~\cite{GSV,Scott},
which is the algebra generated by certain preferred
sets of coordinates on $\Gr(4,n)$.
These coordinates come in two related varieties:
the $\mathcal{A}$-coordinates, which consist of the Pl\"ucker coordinates
and certain homogeneous polynomials in them, and the
$\mathcal{X}$-coordinates~\cite{FG03b},
which consist of certain scale-invariant ratios
of $\mathcal{A}$-coordinates.

In this paper we focus on the cases $n=6,7$, for which the corresponding
cluster algebras have respectively 15, 49 $\mathcal{A}$-coordinates and
15, 385 $\mathcal{X}$-coordinates\footnote{In some applications it is
sensible to count $x$ and $1/x$ separately,
in which case these numbers would be 30, 770.}.  The reader may find
these coordinates tabulated in~\cite{Golden:2013xva}.
Of course, the $\mathcal{X}$-coordinates
are not algebraically independent since
the dimension of $\Conf_n(\mathbb{P}^3)$ is only $3(n-5)$.
A ``cluster'' is a particular choice of $3(n-5)$
cluster $\mathcal{X}$-coordinates in terms of which all others may be
determined by a simple set of rational transformations called mutations.

A still mysterious but apparently important role is played by the
fact that $\Conf_n(\mathbb{P}^3)$ admits a natural Poisson structure, which it
inherits from the Grassmannian~\cite{GSV}.  A characteristic feature of cluster
coordinates is that within each cluster, the $\mathcal{X}$-coordinates
are log-canonical with respect to this Poisson structure, i.e.
\begin{equation}
\{ \log x_i, \log x_j \} =  B_{ij}\,, \qquad i,j = 1, \ldots, 3(n-5)\,,
\end{equation}
where $B$ is an antisymmetric integer-valued matrix (which for $n=6,7$
only takes the values $0,\pm 1$).

We expect all six- and seven-point $L$-loop
scattering amplitudes in planar SYM theory to
be (generalized) polylogarithm functions of uniform transcendental
weight $2L$ whose symbols may be written in terms of the
$\Gr(4,n)$ cluster coordinates.  For the purpose of writing a symbol
alphabet the relevant question is not how many coordinates are algebraically
independent, but how many are multiplicatively independent---we say
that a finite collection $\{y_1, \ldots, y_m\}$ is multiplicatively
independent if there is no collection of integers $\{n_1, \ldots, n_m\}$
such that $\prod y_i^{n_i} = 1$, i.e.~if the
collection $\{\log y_1,\ldots,\log y_m\}$ is linearly independent
over $\mathbb{Z}$.

As mentioned above there are respectively  15 (385) cluster
$\mathcal{X}$-coordinates $x_i$ for $n=6$ ($n=7$), but the corresponding
sets of $\log x_i$ only span spaces of dimension 9 (42).
Choosing bases for these spaces provides a collection
of 9 (42) multiplicatively independent
ratios to serve as symbol alphabets for building
cluster polylogarithm functions.

\subsection{The $\Gr(4,6)$ Cluster Algebra}
\label{subsec:g46}

For six-point amplitudes
the relevant cluster algebra is $\Gr(4,6)$, which is isomorphic to the
$A_3$ cluster algebra.
Its 15 cluster $\mathcal{A}$-coordinates are just the Pl\"ucker coordinates
$\langle ijkl\rangle$.
This algebra has 15 $\mathcal{X}$-coordinates.
In the notation of~\cite{Golden:2014xqa} these are named
$v_i, x^\pm_i$ for $i=1,2,3$ and $e_i$ for $i=1,\ldots,6$.

The reader may find explicit formulas for these as ratios of Pl\"ucker
coordinates in~\cite{Golden:2014xqa}.  Since one of the goals of this
paper is to make contact with the work of Dixon et.~al.~we will instead
provide this information via
the connection to the variables $u,v,w,y_u,y_v,y_w$ used
in~\cite{Dixon:2011nj,Dixon:2011pw,Dixon:2013eka,Dixon:2014voa,Dixon:2014xca,Dixon:2014iba,Dixon:2015iva}.

The three-dimensional kinematic configuration space
$\Conf_6(\mathbb{P}^3)$ may be parameterized in terms of the three
coordinates
\begin{equation}
\label{eq:ydefs}
y_u = \frac{\ket{1236}\ket{1345}\ket{2456}}{\ket{1235}\ket{1246}\ket{3456}}\,, \quad
y_v = \frac{\ket{1235}\ket{1456}\ket{2346}}{\ket{1234}\ket{1356}\ket{2456}}\,,\quad
y_w = \frac{\ket{1246}\ket{1356}\ket{2345}}{\ket{1256}\ket{1345}\ket{2346}}\,.
\end{equation}
Note that a cyclic rotation $Z_i \to Z_{i+1}$ maps
\begin{equation}
y_u \to 1/y_v\,, \qquad y_v \to 1/y_w\,, \qquad y_w \to 1/y_u\,,
\end{equation}
while reflection $Z_i \to Z_{1{-}i}$ (all indices are understood
to be cyclic modulo 6) takes
\begin{equation}
y_u \to y_v\,, \qquad y_v \to y_u\,, \qquad y_w \to y_w\,.
\end{equation}
The spacetime parity operator acts on momentum twistors as\footnote{The notation
means that $W_i$ spans the one-dimensional subspace orthogonal to the
3-plane spanned by $Z_{i-1}, Z_i, Z_{i+1}$ in $\mathbb{C}^4$.}
\begin{equation}
Z_i \to W_i = *(Z_{i-1} \wedge Z_i \wedge Z_{i+1})\,,
\end{equation}
which transforms the cross-ratios defined in~(\ref{eq:ydefs}) according to
\begin{equation}
y_u \to 1/y_u\,, \qquad y_v \to 1/y_v\,, \qquad y_w \to 1/y_w\,.
\end{equation}
It is a curious accident that for $n=6$ spacetime parity reversal is equivalent
on $\Conf_n(\mathbb{P}^3)$ to an element (namely, shift-by-three) of the
cyclic group.

Three other variables used by Dixon et.~al.~may be defined in terms of
these via
\begin{equation}
u = \frac{y_u (1{-}y_v) (1{-}y_w)}{(1{-}y_u y_v)(1{-}y_u y_w)}\,, \quad
v = \frac{y_v (1{-}y_u) (1{-}y_w)}{(1{-}y_u y_v)(1{-}y_v y_w)}\,, \quad
w = \frac{y_w (1{-}y_u) (1{-}y_v)}{(1{-}y_u y_w)(1{-}y_v y_w)}\,.
\end{equation}
Central to our investigations is the Poisson structure
on $\Conf_6(\mathbb{P}^3)$, which may be expressed
in terms of the $y$ variables as
\begin{equation}
\label{eq:pb}
\{ \log y_u, \log y_v \} = \{ \log y_v, \log y_w \} = \{ \log y_w, \log y_u \} = \frac{(1{-}y_u)(1{-}y_v)(1{-}y_w)}{1{-}y_uy_vy_w}.
\end{equation}
It is invariant under the full cyclic group (and hence, it is parity symmetric) but antisymmetric
under reflection.

In terms of these variables, the cluster $\mathcal{X}$-coordinates
may be expressed as
\begin{alignat}{3}
v_1 &= \frac{1-v}{v}\,,
&\qquad v_2 &= \frac{1-w}{w}\,,
&\qquad v_3 &= \frac{1-u}{u}\,,\nonumber \\
x^+_1 &= \frac{y_v (1-y_u y_w)}{1-y_v}\,,
&\qquad x^+_2 &= \frac{y_w (1-y_u y_v)}{1-y_w}\,,
&\qquad x^+_3 &= \frac{y_u (1-y_v y_w)}{1-y_u}\,,\nonumber \\
x^-_1 &= \frac{1-y_u y_w}{y_u y_w (1-y_v)}\,,
&\qquad x^-_2 &= \frac{1-y_u y_v}{y_u y_v (1-y_w)}\,,
&\qquad x^-_3 &= \frac{1-y_v y_w}{y_v y_w (1-y_u)}\,, \\
e_1 &= \frac{1-y_v}{y_v (1-y_u)}\,,
&\qquad e_2 &= \frac{y_v (1-y_w)}{1-y_v}\,,
&\qquad e_3 &= \frac{1-y_u}{y_u (1-y_w)}\,, \nonumber\\
e_4 &= \frac{y_u (1-y_v)}{1-y_u}\,,
&\qquad e_5 &= \frac{1-y_w}{y_w (1-y_v)}\,,
&\qquad e_6 &= \frac{y_w (1-y_u)}{1-y_w}\,.\nonumber
\end{alignat}
Note that under a cyclic shift $Z_i \to Z_{i+1}$ we have
\begin{equation}
\label{eq:cyclic}
v_i \to v_{i+1}\,, \qquad
x_i^\pm \to x_{i+1}^\mp\,, \qquad
e_i \to e_{i+1}\,,
\end{equation}
while under parity
the $v_i$ are invariant and
\begin{equation}
x_i^\pm \to x_i^\mp\,, \qquad
e_i \to e_{i+3}\,.
\end{equation}

Of particular importance are pairs
$x_1,x_2$ of distinct $\mathcal{X}$-coordinates with simple
Poisson brackets.  By ``simple''
we mean specifically that $\{ \log x_1, \log x_2\}$ is either
0 or $\pm 1$.
There are three pairs with Poisson bracket zero,
\begin{equation}
\label{eq:pb0pairs}
\{\log x_i^+, \log x_i^- \} = 0\,,
\end{equation}
and 30 pairs with Poisson bracket $+1$,
\begin{equation}
\label{eq:pb1pairs}
\{\log e_i, \log e_{i+4}\} = \{\log x^\pm_{i+1}, \log v_i\} = \{\log v_{i+1}, \log x^\pm_i\} = \{ \log x_{i+1}^\pm, \log e_i \} = 1
\end{equation}
together with their cyclic images, for $6+6+6+12=30$ pairs.
The remaining 72 pairs have ``complicated'' Poisson brackets (specifically,
non-integer-valued; see
for example~(\ref{eq:pb})).

\subsection{The $\Gr(4,7)$ Cluster Algebra}

For seven-point amplitudes the relevant cluster is algebra is
$\Gr(4,7)$, which is isomorphic to the $E_6$ algebra.
The 49 cluster $\mathcal{A}$-coordinates consist of the
35 Pl\"ucker coordinates $\ket{ijkl}$ together with 14
homogeneous polynomials denoted by $\ket{1(23)(45)(67)}$,
$\ket{2(13)(45)(67)}$ (and their cyclic images), where
\begin{equation}
\ket{ i( i{-}1,i{+}1)(j,j{+}1)(k,k{+}1)}
=\ket{i{-}1\,i\,j\,j{+}1} \ket{i\,i{+}1\,k\,k{+}1}-
\ket{i{-}1\, i\ k\, k{+}1} \ket{i\,i{+}1\,j\,j{+}1}\,.
\end{equation}
One can build from these 49 $\mathcal{A}$-coordinates a total of 385
cluster $\mathcal{X}$-coordinates (or 770 if we count
their multiplicative inverses).
These are tabulated on pages 40--41 of~\cite{Golden:2013xva}.
Out of $\frac{1}{2} \cdot 385 \cdot 384 = 73920$ pairs of
$\mathcal{X}$-coordinates, 2520 have Poisson bracket $\pm 1$ while
833 have Poisson bracket zero.

\subsection{The Cobracket and Bloch Groups}

We recall that the algebra $\mathcal{A}$ of generalized polylogarithm
functions admits a coproduct giving it
the structure of a Hopf algebra~\cite{G02}.
When we work with the quotient space $\mathcal{L}$ of
polylogarithm functions modulo products of functions of lower weight,
the coproduct descends onto the quotient space
to a cobracket $\delta$ which satisfies $\delta^2=0$.
We review here only the barest essentials, and refer
the reader to~\cite{Golden:2013xva,Golden:2014xqa} for additional details.

The cobracket of a weight-4 function has two components,
\begin{equation}
\delta \mathcal{L}_4 \in (B_3 \otimes \mathbb{C}^*) \oplus (B_2
\wedge B_2)\,,
\end{equation}
where the Bloch group $B_k$ is, for our purposes, the free abelian
group generated by functions of the form
$\{x\}_k \equiv - \Li_k(-x)$, where $\Li_k$ is the classical
polylogarithm function and $x$ is a function on $\Conf_n(\mathbb{P}^3)$
which is rational in Pl\"ucker coordinates.

The fact that $\delta^2 = 0$ and that $\delta$ has trivial cohomology
means that if $a \in B_3 \otimes \mathbb{C}^*$ and $b \in B_2 \wedge B_2$,
then there exists a function $f$ whose cobracket components are $a \oplus
b$ if and only if $\delta_{31}(a) + \delta_{22}(b) = 0$.  As explained
in~\cite{Golden:2014xqa}, this condition can be used to explicitly enumerate
cluster functions, at least on algebras of finite type.  For
such algebras $B_3 \otimes \mathbb{C}^*$ and $B_2 \wedge B_2$ are
finite dimensional vector spaces on which $\delta$ acts linearly, so
the space of cluster $\mathcal{A}$-functions is simply
the kernel of $\delta$.

At weight 4 a general polylogarithm can be expressed in terms of the
classical functions $\Li_k$ if and only if its $B_2 \wedge B_2$
cobracket component vanishes.
We will often be interested in counting the number of
non-classical functions, since the classical ones (which correspond
to solutions of $\delta_{31}(a) = 0$) are trivial to enumerate.
To answer this question we compute the dimension of the subspace
of $B_2 \wedge B_2$ such that the equation
$\delta_{31}(a) + \delta_{22}(b) = 0$
is solvable for some $a \in B_3 \otimes \mathbb{C}^*$.

One final piece of terminology concerns the interplay between
the Poisson structure on the Grassmannian cluster algebras and the
cobracket of polylogarithm functions.  We recall that two
cluster $\mathcal{X}$-coordinates
$x, y$ have $\{ \log x, \log y \} \in \mathbb{Z}$
only if there exists a cluster containing either $x$ or $1/x$, and either
$y$ or $1/y$.  As reviewed in~\cite{Golden:2013xva}, the combinatorics
of mutations is encoded in a graph called the (generalized)
Stasheff polytope associated to the algebra.  We therefore
say that a function has ``Stasheff local'' $B_2 \wedge B_2$ if
it can be expressed as a linear combination of terms of the form
$\{ x \}_2 \wedge \{ y \}_2$ for pairs having integer
Poisson bracket (for $\Gr(4,6)$
and $\Gr(4,7)$, this integer will always be in the set $\{-1,0,+1\}$).

\section{The Cluster Structure of Hexagon Functions at Weight 4}

\subsection{Setup}

In this section we consider cluster functions on the $A_3 \cong \Gr(4,6)$
cluster algebra.
The term ``cluster $\mathcal{A}$-function''
introduced in~\cite{Golden:2014xqa} refers, in the present
application, to an integrable
symbol written in the 9-letter alphabet of
cluster coordinates
(specifically, this means any
multiplicatively independent set of $\mathcal{X}$-coordinates; or equivalently,
homogeneous ratios of
$\mathcal{A}$-coordinates)
on $\Gr(4,6)$.

Any linear combination of cluster $\mathcal{A}$-functions with the
property that only the three variables $u, v, w$ appear in the first-entry
of the symbol, reflecting the physically allowed branch points for
a scattering amplitude~\cite{Gaiotto:2011dt},
is called a ``physical function'' or, following
the terminology of~\cite{Dixon:2013eka}, a ``hexagon function''.
These have been studied through high weight
in the series of papers~\cite{Dixon:2011nj,Dixon:2011pw,Dixon:2013eka,Dixon:2014voa,Dixon:2014xca,Dixon:2014iba,Dixon:2015iva},
but we restrict our analysis to weight 4 as our aim is to explore
connections between the cobrackets and the cluster Poisson structure
of these
functions.

Let $\mathcal{A}_k$ denote the vector space of all
weight-$k$ cluster $\mathcal{A}$-functions.
Such functions are easy to count for any $A_m$ type cluster algebra
(see~\cite{Brown,Parker:2015cia}); for $A_3$ we have the generating
function
\begin{equation}
\label{eq:gf}
f_{A_3}(t) = 1 + \sum_{k=1}^\infty t^k \dim(\mathcal{A}_k)
= \frac{1}{1-2 t} \frac{1}{1 - 3t} \frac{1}{1 - 4t}\,,
\end{equation}
so that
\begin{equation}
\dim(\mathcal{A}_k) = 9, 55, 285, 1351, \ldots \qquad k=1,2,3,4,\ldots\,.
\end{equation}
Let $\mathcal{L}_k$ denote the quotient of $\mathcal{A}_k$ by products of
functions of lower weight.
The number of such functions can be computed by taking the plethystic
logarithm of the generating
function $f_{A_3}(t)$ (see for example~\cite{Benvenuti:2006qr}), which
gives
\begin{equation}
\label{eq:dimlk}
\dim(\mathcal{L}_k) = 9, 10, 30, 81, \ldots \qquad k=1,2,3,4,\ldots\,.
\end{equation}
Finally we denote by $B_k$ the subspace of $\mathcal{L}_k$ generated by the
classical polylogarithms (we do not yet restrict their arguments
to be cluster $\mathcal{X}$-coordinates).  We have
\begin{equation}
\label{eq:dimbk}
\dim(\mathcal{B}_k) = 10, 30, 45, \ldots \qquad k=2,3,4,\ldots\,.
\end{equation}
For $k<4$ the agreement with~(\ref{eq:dimlk})
reflects the fact that all such generalized polylogarithms
can be expressed in terms of the classical functions; for higher
$k$ these numbers can be obtained by choosing a basis for
$\mathcal{L}_k$ and computing $\dim \ker \delta$ as described
in the previous section.

\subsection{The Non-Classical Functions}

\label{sec:nc6}

Beginning at $k=4$ we can distinguish between classical and non-classical functions.
At weight $k=4$, the ``non-classicalness'' of a function is completely
characterized by its $B_2 \wedge B_2$ cobracket
component (see for example~\cite{Golden:2013xva}).
Since $B_2$ has dimension 10 according to~(\ref{eq:dimbk}),
$B_2 \wedge B_2$ evidently has dimension 45.
However, a random element of this vector space is not guaranteed to
be the $B_2 \wedge B_2$ cobracket component of
any cluster
$\mathcal{A}$-function---there is a nontrivial integrability constraint.

In fact, by comparing~(\ref{eq:dimbk})
to~(\ref{eq:dimlk}) we see that there are 81 functions in all, minus
45 classical functions, for a total of 36 non-classical functions.
We conclude that in the 45-dimensional space $B_2 \wedge B_2$ spanned
by objects of the form $\{x\}_2 \wedge \{y\}_2$, for cluster
coordinates $x$ and $y$, only the linear combinations lying in
a particular 36-dimensional subspace correspond to cobracket components
of actual
cluster $\mathcal{A}$-functions.\footnote{Linear combinations which fall
outside this 36-dimensional subspace are
certainly integrable~\cite{Galois},
but they integrate to functions with symbols involving letters
which are not cluster coordinates, for example differences of
$\mathcal{X}$-coordinates
$x_i - x_j$, which does not in general factor
into a product of cluster coordinates.  Hence they are not
cluster $\mathcal{A}$-functions.}
We will shortly characterize this 36-dimensional space completely.

Let us write $PB_0$ to denote the subspace of
$B_2 \wedge B_2$ spanned by objects of the form $\{x\}_2 \wedge
\{y\}_2$ for pairs having Poisson bracket $\{ \log x, \log y\} = 0$.
In what follows we will for example say that a function ``lives
in $PB_0$'' if its $B_2 \wedge B_2$ cobracket component can
be expressed in terms of such pairs.
Similarly, let $PB_1$ be the subspace spanned by pairs having
Poisson bracket $1$, and let us also use the shorthand
$PB_* = B_2 \wedge B_2$, meaning that the Poisson bracket can be
anything.
We found in~(\ref{eq:pb0pairs}) and~(\ref{eq:pb1pairs})
that there are respectively 3, 30 pairs with Poisson bracket
0, $1$.  It is simple to check that the corresponding elements
are linearly independent in $B_2 \wedge B_2$, so we have that
$\dim PB_0 = 3$
and $\dim PB_1 = 30$, while of course
$\dim PB_* = \dim B_2 \wedge B_2 = 45$.

With this notation in hand let us now summarize our findings on
the 36 non-classical cluster $\mathcal{A}$-functions
at weight four, which we find
fall into two broad groups:

(A) 6 of these functions are the
``$A_2$ cluster functions'' introduced in~\cite{Golden:2014xqa}.
There is one such function for each $A_2$ subalgebra of $A_3$;
these subalgebras and the associated functions are represented
visually in equation~(4.3) of that paper.
These six functions
have additional ``cluster structure'':
their $B_3 \otimes \mathbb{C}^*$ cobracket components
can be expressed entirely in terms of cluster
$\mathcal{X}$-coordinates---this means that they are ``cluster
$\mathcal{X}$-functions'' in the terminology of~\cite{Golden:2014xqa}.
General elements of this six-dimensional space are not
Stasheff local---their
$B_2 \wedge B_2$ cobracket components are not expressible
in terms of pairs of coordinates with Poisson bracket $0, \pm 1$.
Only one particular linear combination of these 6---the one called the
$A_3$ function in~\cite{Golden:2014xqa}---has
a nice $B_2 \wedge B_2$,
in fact lying inside $PB_0$.  The $B_2 \wedge B_2$ cobracket component
of this $A_3$ function is
\begin{equation}
\label{eq:one}
\sum_{i=1}^3 \{ x^+_i \}_2 \wedge \{ x^-_i \}_2\,.
\end{equation}
This quantity is parity-odd so it cannot possibly appear in the
two-loop six-point
MHV remainder function, which is parity-even.  This ``explains''
why the hypothesis that two-loop MHV remainder functions
must live in $PB_0$,
which we know to be true for all $n$~\cite{Golden:2014pua},
implies that the case $n=6$ must be classical.

(B) The remaining 30 functions are sort of the opposite:
no linear combination
of these 30 has a $B_3 \otimes
\mathbb{C}^*$ content which can be expressed entirely in terms of
$\mathcal{X}$-coordinates, so none
of them are cluster
$\mathcal{X}$-functions.  On the other hand, all
of them are Stasheff local---they all have ``nice''
$B_2 \wedge B_2$, in fact they span exactly the
30-dimensional subspace $PB_1 \subset B_2 \wedge B_2$.

\subsection{The Physical (Hexagon) Functions}

Dixon et.~al.~find that there are precisely
15 functions at weight 4
(modulo products of functions of lower weight)
satisfying the first-entry condition, which they call hexagon
functions.  Let us put aside 9 which are purely classical and focus
on the two types of functions named $\Omega_2$ and
$F_1$
in~\cite{Dixon:2013eka}.

(A) The function $F_1$ is parity-odd and comes in three cyclic permutations
(i.e., $i \to i{+}2$ and $i \to i{+}4$).
These functions are rather interesting; each of them has a $B_2 \wedge
B_2$ coproduct component given by~(\ref{eq:one}) plus additional terms
which cannot be expressed in terms of pairs having simple Poisson bracket.
Since~(\ref{eq:one}) is invariant under $i \to i{+}2$, we can throw
out these terms by taking the difference between any two pairs of the
three permutations of $F_1$.  Indeed such linear combinations have
appeared in the literature, as in (B.18) and (B.20) of~\cite{Dixon:2013eka}
which define the function $\widetilde{V}$ by
\begin{equation}
8 \widetilde{V}=
- F_1(u,v,w) + F_1(w,u,v) + \text{products of lower-weight functions}.
\end{equation}
Hence only two of the three distinct cyclic permutations of
$\widetilde{V}$ are linearly independent.

(B) Next we look at the parity-even function $\Omega_2$ which also
comes in three cyclic permutations.  At the level of $B_2 \wedge B_2$,
where we can ignore all terms involving only classical polylogarithms,
the function $\Omega_2$ is equivalent (modulo an overall multiplicative
factor) to the function called $V$ by Dixon et.~al.; see
for example (7.1) through (7.3) of~\cite{Dixon:2011nj}.
In that paper it was also observed that the three cyclic permutations
of this function add up to a purely classical function, so the
three different permutations of $V$ span only a two-dimensional subset
of $B_2 \wedge B_2$.

To summarize, we find that the subspace of $B_2 \wedge B_2$ spanned
by physical (hexagon) functions has dimension 5.
Two dimensions are spanned by the parity-even functions of type $V$,
while three dimensions are spanned by the parity-odd functions of
type $F_1$.  Although a generic vector in the three-dimensional
parity-odd subspace has terms with ``bad'' Poisson brackets, there
is something especially nice about the
subspace spanned by the permutations of
$V$ and $\widetilde{V}$ together.  To see this we exhibit here
a formula for their cobracket components, which we find are most
simply packaged in the formula
\begin{equation}
\label{eq:a}
\begin{aligned}
\delta\rvert_{2,2}(V + \widetilde{V}) = \frac{1}{2} \{v_2\}_2 \wedge \{x^-_1\}_2 - \frac{1}{2} \{v_1\}_2 \wedge \{x^-_3\}_2 - \frac{1}{2} \{x^+_1\}_2 \wedge \{v_3\}_2 + \frac{1}{2} \{x^+_2\}_2 \wedge \{v_1\}_2.
\end{aligned}
\end{equation}
Since $V$, $\widetilde{V}$ have parity even and odd, respectively,
$\delta\rvert_{2,2}(V - \widetilde{V})$ is
given by the same formula but with $x^\pm \to x^\mp$.
We now see that each term in~(\ref{eq:a})
involves only the $PB_1$ pairs listed in~(\ref{eq:pb1pairs})!
Moreover, it is trivial to check directly from~(\ref{eq:a}) and the 
cyclic transformations~(\ref{eq:cyclic}) that the six functions
$V$, $\widetilde{V}$ altogether span only a four-dimensional subspace
of $PB_1$.

\subsection{Summary}
\label{sec:summary6}

The results of this section can be summarized in the following
classification of weight-4 cluster functions on $A_3 \cong \Gr(4,6)$:
\begin{align*}
&\text{There are a total of}~81~\text{irreducible weight-four cluster $\mathcal{A}$-functions} \\
&\qquad \rotatebox[origin=c]{180}{$\Lsh$}~45~\text{classical, 10 of which are physical} \\
&\qquad \rotatebox[origin=c]{180}{$\Lsh$}~36~\text{non-classical, 5 of which are physical (three permutations of $F_1$ and two of $\Omega_2$)} \\
&\qquad \qquad \rotatebox[origin=c]{180}{$\Lsh$}~30\ PB_1~\text{functions, 4 of which are physical (two permutations each of $V, \tilde{V}$)} \\
&\qquad \qquad \rotatebox[origin=c]{180}{$\Lsh$}~6\ A_2~\text{functions;
these are all of the cluster $\mathcal{X}$-functions} \\
&\qquad \qquad \qquad \rotatebox[origin=c]{180}{$\Lsh$}~1\ PB_0~\text{function, the $A_3$ function} \\
&\qquad \qquad \qquad \rotatebox[origin=c]{180}{$\Lsh$}~5\ PB_*~\text{functions}
\end{align*}
Let us emphasize that these numbers count only irreducible
functions, and that starting from the third line they moreover
count functions modulo the classical function $\Li_4$ (i.e., the numbers refer
to dimensions of subspaces of $B_2 \wedge B_2$).
When we say that a function is physical modulo additional terms, we mean
that it is possible to choose the additional terms to render the function
physical.

\subsection{The Two-Loop Hexagon MHV Amplitude}

Let us now comment on the relevance of these functions
to the two-loop six-point MHV remainder function $R_6^{(2)}$, which was found
to be expressible in terms of the classical polylogarithm
functions $\Li_k$ in~\cite{Goncharov:2010jf} (a fact that we ``explained''
below~(\ref{eq:one})).
In fact, this amplitude is even more special because it is
a cluster $\mathcal{X}$-function, which means that it
can be expressed in entirely in terms of the $\Li_k(-x)$; the
$\Li_k(1+x)$ and $\Li_k(1+1/x)$ functions, whose
$B_3 \otimes \mathbb{C}^*$
cobracket
components are not expressible in terms of cluster $\mathcal{X}$-coordinates,
are not needed~\cite{Golden:2013xva}.

Above we tabulated our finding that (modulo products of lower-weight
functions) there are only 10 physical and classical polylogarithms
at weight four.
In this space we now search for functions whose coproducts
are expressible entirely in terms of the $\Li_k(-x)$.
We find that there is a unique
linear combination that is invariant under the discrete symmetries
(parity and dihedral invariance) that MHV amplitudes must possess.
That linear combination is proportional to the two-loop MHV remainder
function
\begin{equation}
R_6^{(2)\,\rm MHV} = \sum_{i=1}^3 \left[ \Li_4(-x_i^+) + \Li_4(-x_i^-)
- \frac{1}{2} \Li_4(-v_i)\right] + \text{products
of lower-weight functions,}
\end{equation}
in agreement with the known result~\cite{Goncharov:2010jf}.
(This argument, of course, does not fix the overall coefficient.)
Of course, in this case it is very well known that the product
terms are also completely fixed by simple considerations, but our focus
in this paper is on the leading term.

\subsection{The Two-Loop Hexagon NMHV Amplitude}

The $n=6$ NMHV two-loop ratio function is given by~\cite{Dixon:2011nj}
\begin{equation}
\mathcal{P}_{6,{\rm NMHV}}^{(2)} = [23456]  [V(u,v,w) + \widetilde{V}(y_{u},y_{v},y_{w})] + \text{cyclic}
\label{eq:nmhv6}
\end{equation}
where $[23456]$ is the $R$-invariant
\begin{equation}
\label{eq:Rinvariant}
[abcde] = \frac{\delta^{4}\left(\chi_{a}\langle bcde \rangle + cyclic \right)}{\langle abcd \rangle \langle bcde \rangle \langle cdea \rangle \langle deab\rangle \langle eabc\rangle}
\end{equation}
and
$V$, $\widetilde{V}$ are the two generalized polylogarithm functions of
uniform transcendental weight four reviewed in Section 3.3 above.
These two functions were computed
explicitly in~\cite{Dixon:2011nj} (see also~\cite{Parker:2015cia} for
a different presentation of these functions).
The $B_2 \wedge B_2$ component of the cobracket of this amplitude was
computed in~(\ref{eq:a}), where it was found to be expressible
entirely in terms of pairs
living in $PB_1$\footnote{This observation was first made by
C.~Vergu~\cite{Vergu}.}.

The NMHV ratio function provides us
(at the level of $B_2 \wedge B_2$) with a total of four linearly
independent
non-classical functions of weight 4 (as reviewed above,
each of $V$ and $\widetilde{V}$ comes in three cyclic permutations,
but the cyclic sum of each is separately zero inside $B_2 \wedge B_2$).
We see from the summary in Section~\ref{sec:summary6}
that precisely 5 functions of this type exist.
Only four linear combinations of them, however, actually
appear in the amplitude---these are precisely the four linear
combinations which live in $PB_1$!
The one additional
non-classical
weight-4 hexagon function which exists but does
not appear in the amplitude, $F_1$ by itself, has terms
with ``bad'' Poisson brackets (i.e., non-Stasheff local terms)
in its $B_2 \wedge B_2$ content.

\section{The Cluster Structure of Heptagon Functions at Weight 4}

\subsection{Setup}

In this section the term ``cluster function'' refers to an integrable
symbol written in the 42-letter alphabet of cluster coordinates
on $\Gr(4,7)$.  Any linear combination of such symbols with the
property that only the Pl\"ucker coordinates
of the form
$\ket{i\,i{+}1\,j\,j{+}1}$ appear in the first entry
of the symbol, reflecting the physically allowed branch points for
a scattering amplitude, is called (the symbol of) a
``physical function'' or
a ``heptagon function'' following
the terminology of~\cite{Drummond:2014ffa} where they have been studied
through weight six.
The analysis here, where we aim to
make finer statements about the connection to the Poisson bracket
of the cluster algebra, is again restricted to weight 4, of relevance to
two-loop amplitudes.

Let $\mathcal{A}_k$ denote the vector space of all weight-$k$
functions.  In contrast to the $A_m$ cluster algebras and the
example shown in~(\ref{eq:gf}), we do not know of any generating
function which counts the number of cluster functions for the $E_6$
algebra.
These may be tabulated through weight 3 by explicit enumeration,
but at higher weight these numbers must be computed by analyzing
the integrability constraint.  This boils down to a linear algebra
problem, since counting the number of cluster functions at weight $k$
is the same as finding how many linear combinations of the
$42^k$ weight-$k$ symbols satisfy the integrability constraint.
(This calculation can be rendered more manageable by imposing
integrability at the level of the cobracket rather than at the level of
the symbol.)
We have carried this out at $k=4$ to find that
\begin{equation}
\dim(\mathcal{A}_k) = 42, 1035, 19536, 312578, \ldots \qquad k=1,2,3,4,\ldots\,.
\end{equation}
Let $\mathcal{L}_k$ denote the quotient of $\mathcal{A}_k$ by products of
functions of lower weight.  As in~(\ref{eq:dimlk}) taking the
plethystic logarithm~\cite{Benvenuti:2006qr} gives
\begin{equation}
\label{eq:dimlk7}
\dim(\mathcal{L}_k) = 42, 132, 748, 4193, \ldots \qquad k=1,2,3,4,\ldots\,.
\end{equation}
Finally we denote by $B_k$ the subspace of $\mathcal{L}_k$ generated by the
classical polylogarithms (we do not yet restrict their arguments
to be cluster $\mathcal{X}$-coordinates).  We have
\begin{equation}
\label{eq:dimbk7}
\dim(\mathcal{B}_k) = 132, 748, 1155, \ldots \qquad k=2,3,4,\ldots\,.
\end{equation}
As mentioned before, agreement
of these numbers with~(\ref{eq:dimlk7})
is guaranteed for $k<4$, and we obtained the value 1155 for $k=4$ by
computing $\dim \ker \delta$ as described in Section~2.

Before we turn to weight 4, a minor interesting comment about $k=3$
is in order.
It is simple to write down classical cluster functions of the form
$\Li_k(-x)$, $\Li_k(1+x)$ and $\Li_k(1+1/x)$
for any weight $k$, where $x$ runs over
the set of 385 $\mathcal{X}$-coordinates.
For $k=3$, this set of functions is overcomplete due to the identity
\begin{equation}
\Li_3(-x) + \Li_3(1+x) + \Li_3(1+1/x) = 0 \quad
\text{mod products
of lower-weight functions.}
\end{equation}
Among the 385 functions of type $\Li_3(-x)$ there are exactly
22 additional linear relations. These were discovered
in~\cite{Golden:2013xva}, where they were called $D_4$ identities
since the simplest manifestation of this identity occurs for the $D_4$
algebra.  Altogether then these identities account for the
$3 \times 385 - 385 - 22 = 748$ linearly independent weight-3
cluster $\mathcal{A}$-functions tabulated in~(\ref{eq:dimlk7}).

\subsection{The Non-Classical Functions}

Let us now repeat the analysis done in
the beginning of Section~\ref{sec:nc6}
for the $E_6$ algebra.
Since $B_2$ has dimension 132, $B_2 \wedge B_2$ has dimension
8646.
We again use the notation $PB_0$, $PB_1$, and $PB_* = B_2 \wedge B_2$
to denote
the subspaces
spanned by elements of the form $\{x\}_2 \wedge \{y\}_2$
for pairs $x,y$ having Poisson bracket 0, $\pm 1$,
or ``anything.''
We find that $PB_0$ has dimension 455 and $PB_1$ has dimension 2520.

A quick glance at~(\ref{eq:dimlk7}) and~(\ref{eq:dimbk7})
reveals that there are $4193 - 1155 = 3038$ non-classical cluster
functions at weight $k=4$.
We find that these fall into three groups:

(A) First, there are the $A_2$ functions.
We recall
from (for example)~\cite{Golden:2013xva} that $E_6$ has 1071
$A_2$ subalgebras, so one can construct 1071 $A_2$ functions
according to the definition given in~\cite{Golden:2014xqa}, but
only 448 of these are linearly independent inside
$B_2 \wedge B_2$\footnote{This result
was first obtained in the undergraduate thesis of A.~Scherlis.}.
These functions are moreover cluster $\mathcal{X}$-functions:
their $B_3 \otimes \mathbb{C}^*$ cobracket components
can be expressed entirely in terms of cluster $\mathcal{X}$-coordinates,
but their $B_2 \wedge B_2$ content is, in general,
not Stasheff local---not
expressible in terms of pairs with Poisson bracket $0, \pm 1$.

There are no linear combinations of these 448 functions which
live in $PB_1$---these are covered in (B) just ahead---but
we find that 195 linear combinations live in $PB_0$.  This 195-dimensional
space is spanned by the set of $A_3$ functions associated to the
various $A_3$ subalgebras of $E_6$.

(B) There are 2520 functions which span the 2520-dimensional
subspace
$PB_1 \subset B_2 \wedge B_2$.
We found the same phenomenon in the six-point case discussed in the
previous section.  There we furthermore found that no linear combination of
these $PB_1$ functions had a $B_3 \otimes \mathbb{C}^*$ component that
could be expressed entirely in terms of $\mathcal{X}$-coordinates.
We have not repeated this analysis for the 2520 seven-point functions;
the computation seems formidable.

(C) There are an additional $3038 - 448 - 2520 = 70$ functions which
we can tabulate explicitly (at least at the level of their cobrackets),
but seem to have no nice characterization.

\subsection{The Physical (Heptagon) Functions}

It was found in~\cite{Drummond:2014ffa} that
there are precisely $1288$ functions at weight 4
satisfying the first-entry condition, which are called physical,
or heptagon functions.
We have computed the $B_2 \wedge B_2$ cobracket of each of them,
and found that there are only 126 non-zero linear combinations.
This means that there are 1162 classical heptagon functions
and 126 non-classical heptagon functions at weight 4.
We have found that these 126 heptagon functions fall into three types:

(A) A total of $105$
of these functions live in $PB_0$; they come in $15$ families related
by cyclic permutations.

(B) A total of 14 of these functions live in $PB_1$;
they come in 2 families related by cyclic permutations.

(C) There is one remaining family  of 7 functions related by cyclic
permutations. No linear combination of these
is Stasheff local (i.e.,
lives within the union
of $PB_0$ and $PB_1$).

\subsection{Summary}
\label{sec:summary7}

The results of this section can be summarized in the following
classification of weight-4 cluster functions on $E_6 \cong \Gr(4,7)$:
\begin{align*}
&\text{There are a total of}~4193~\text{irreducible weight-four cluster $\mathcal{A}$-functions} \\
&\qquad \rotatebox[origin=c]{180}{$\Lsh$}~1155~\text{classical, 770 of which are physical} \\
&\qquad \rotatebox[origin=c]{180}{$\Lsh$}~3038~\text{non-classical, 126 of which are physical} \\
&\qquad \qquad \rotatebox[origin=c]{180}{$\Lsh$}~2520\ PB_1~\text{functions, 105 of which are physical} \\
&\qquad \qquad \rotatebox[origin=c]{180}{$\Lsh$}~448\ A_2~\text{functions;
these are all of the cluster $\mathcal{X}$-functions} \\
&\qquad \qquad \qquad \rotatebox[origin=c]{180}{$\Lsh$}~195\ PB_0~\text{function, 14 of which are physical} \\
&\qquad \qquad \qquad \rotatebox[origin=c]{180}{$\Lsh$}~253\ PB_*~\text{functions} \\
&\qquad \qquad \rotatebox[origin=c]{180}{$\Lsh$}~70~\text{other}~PB_*~\text{functions}
\end{align*}
Again let us emphasize that these numbers count only irreducible
functions, and that starting from the third line they moreover
count functions modulo the classical function $\Li_4$ (i.e., the numbers refer
to dimensions of subspaces of $B_2 \wedge B_2$).
When we say that a function is physical modulo additional terms, we mean
that it is possible to choose the additional terms to render the function
physical.

\subsection{The Two-Loop Heptagon MHV Amplitude}

The symbol of the two-loop seven-point MHV remainder function
$R_7^{(2)}$
was computed in~\cite{CaronHuot:2011ky}, and
its cobracket was computed in~\cite{Golden:2013xva}, where
it was observed to be a cluster $\mathcal{X}$-function
living in $PB_0$.  An analytic formula
for $R_7^{(2)}$ was obtained in~\cite{Golden:2014xqf} and checked
against the earlier numerical results
of~\cite{Anastasiou:2009kna}.

If we start from the hypothesis that $R_7^{(2)}$ should
be a cluster $\mathcal{X}$-function living in $PB_0$, then we see from
the above chart that there are only 14 physical functions with
these properties.
It was shown in~\cite{Golden:2014pua} that
only one linear combination of these
has the dihedral symmetry required of the amplitude, is well-defined
in the collinear limit, and satisfies the ``last-entry''
condition~\cite{CaronHuot:2011ky} required by supersymmetry.

In fact these constraints, while all true, are vastly stronger
than necessary to pin down $R_7^{(2)}$:
in~\cite{Drummond:2014ffa} it was found that the symbol of
$R_7^{(2)}$ is the unique weight-4 heptagon function (up to an
overall multiplicative factor) which is well-defined in all
$i{+}1 \parallel i$ collinear limits!

\subsection{The Two-Loop Heptagon NMHV Amplitude}

The symbol of the seven-point 2-loop NMHV ratio function
$\mathcal{P}_{7,{\rm NMHV}}^{(2)}$ was first computed
in~\cite{CaronHuot:2011kk}.
It may be expressed as a linear combination of the 21 seven-point
NMHV
$R$-invariants (of which 15 are linearly independent), with
coefficients that have uniform transcendentality weight 4.
Due to the linear relations between $R$-invariants there is
some freedom in how to represent the amplitude (i.e., one can
shift terms from one transcendental function to another by
adding zero to the amplitude in various ways).

Despite this freedom,
we find
that it impossible to write the $B_2 \wedge B_2$ cobracket
of this amplitude
in a Stasheff local manner, i.e.~in terms of
$\{x\}_2 \wedge \{y\}_2$ for pairs $x, y$
having Poisson bracket $0, \pm 1$.
The local terms having ``good'' Poisson brackets may be expressed (in
one particular representation of the amplitude) as
\begin{equation}
\label{eq:nmhv7}
\delta_{22} \mathcal{P}_{7,{\rm NMHV}}^{(2)}\rvert_{\rm ``good"}
= (f_{12} R_{12} + f_{13} R_{13} + f_{14} R_{14}) +
{\rm cyclic},
\end{equation}
where the quantities $f_{12}$, $f_{13}$ and $f_{13}$
are presented explicitly in the appendix,
and $R_{ij}$ is the
$R$-invariant whose arguments are 1234567 (in that
order) but with
$i$ and $j$ omitted---this
is the same as the notation used in~\cite{Dixon:2011nj}.
Meanwhile the ``bad'' terms are given by:
\begin{equation}
\label{eq:bad}
\delta_{22} \mathcal{P}_{7,{\rm NMHV}}^{(2)}\rvert_{\rm ``bad"} = \left(R_{25}-R_{26}+R_{37}-R_{47}\right) B_1 +\text{cyclic}
\end{equation}
in terms of a single
element $B_1 \in B_2 \wedge B_2$
(also given in the appendix)
which is not expressible solely in terms of pairs
having Poisson bracket zero or one.

In fact we can point our finger directly at the ``offending'' function
corresponding to $B_1$ in the summary presented at the end
of Section~\ref{sec:summary7}.
There we found that of the 126 non-classical weight-4 heptagon functions,
105 live in $PB_1$ while 14 live in $PB_0$, leaving
$127 - 105 - 14 = 7$ unaccounted for.  These other seven functions
have $B_2 \wedge B_2$ cobracket components given exactly by $B_1$
in its seven cyclic arrangements.

\section{Conclusion}

In this paper we have studied in detail the taxonomy of weight-4
cluster functions on the cluster algebras relevant for 6-
and seven-point amplitudes in planar SYM theory.  In particular we
have counted the numbers of linearly independent functions satisfying
various mathematical constraints on their cobrackets, and the physical
``first-entry'' constraint which specifies the locations where
amplitudes are permitted to have branch points on the principal sheet.
These results are summarized in Sections~\ref{sec:summary6}
and~\ref{sec:summary7}.

For $n=6$ the story is very simple:
there is no non-classical weight-4 generalized polylogarithm
function which is consistent with the discrete symmetries of the MHV
amplitude and
whose $B_2 \wedge B_2$ cobracket component is expressible in terms of
pairs of cluster $\mathcal{X}$-coordinates having Poisson bracket 0.
This ``explains'' why the two-loop six-point MHV remainder function ``must
be'' expressible in terms of classical polylogarithms~\cite{Goncharov:2010jf}.

Meanwhile, there are precisely 4 linearly independent non-classical functions
which satisfy the first-entry condition and
are Stasheff local (they have $B_2 \wedge B_2$ cobracket
components are expressible in terms of pairs of
cluster $\mathcal{X}$-coordinates having Poisson bracket 1).
These are
precisely the (non-classical
parts of the) 4 independent
functions which appear in the two-loop six-point NMHV
ratio function~\cite{Dixon:2011nj}.

For $n=7$, as has already been observed
in~\cite{Golden:2014pua,Drummond:2014ffa}, the cobracket
(indeed, the whole symbol) of the two-loop MHV amplitude
is uniquely determined by a simple list of mathematical and physical
constraints.
However the story for the two-loop NMHV ratio function is a little more
complicated.  We find that the cobracket of this amplitude is not
expressible in a Stasheff local manner (that means, in terms of pairs
having Poisson bracket $0, \pm 1$).
It would be very interesting to learn if there is some other question
one may ask about the cluster structure of this amplitude, to which
a more affirmative answer may be given.
We expect to be the case since it is known that there is
a cluster structure at the level of the integrand (aspects of which
have been explored in~\cite{ArkaniHamed:2012nw,Paulos:2014dja}), of
which some echo ought to remain for integrated amplitudes.

One of our results might be of more
mathematical than physical interest.  For both the $A_3$ and $E_6$
cluster algebras, we find that for any pair of $\mathcal{X}$-coordinates
with Poisson bracket $\{ \log x, \log y \} = 1$, there exists
a weight-4 cluster $\mathcal{A}$-function (that is, an integrable
symbol whose letters are drawn from the alphabet of cluster coordinates)
whose $B_2 \wedge B_2$ cobracket component is
$\{ x \}_2 \wedge \{ y \}_2$.  It would be interesting to learn if there
is a mathematical explanation for this fact, and whether it is
valid for more general cluster algebras (in particular, for ones
of infinite type).
In contrast, pairs of $\mathcal{X}$-coordinates having Poisson bracket
0 are rarely integrable in this manner; the two-loop MHV amplitudes
of planar SYM theory remarkably provide functions of this relatively
rare type.

In the introduction we mentioned that in previous work
including~\cite{Golden:2014pua}
it has been remarked that the mathematical and
physical constraints on MHV amplitudes seem almost orthogonal.
This is both good and bad.
On the one hand it is good to discover a short
list of simple criteria which uniquely, or almost uniquely,
determine an amplitude of interest---this is the core
goal of the $S$-matrix program.
On the other hand it is bad when there is no
known formalism which simultaneously manifests both
types of constraints.
We do not yet know of any way, besides explicit enumeration,
to actually identify and
write down functions satisfying both the physical and mathematical
we expect amplitudes to possess.
Explicit results for higher loop planar SYM amplitudes remain, at least
for the moment, difficult needles to find.

\acknowledgments

We have benefitted from stimulating discussions, correspondence,
and collaboration with
James Drummond, John Golden,
Alexander Goncharov,
Daniel Parker,
Adam Scherlis,
Cristian Vergu, and Anastasia Volovich.
We are especially grateful to Lance Dixon for careful comments on the draft.
MS is grateful to the CERN theory group for hospitality during the course
of this work, which
was supported by the US Department of Energy under contract
DE-SC0010010 Task A.

\appendix
\section{Two-Loop Heptagon NMHV Coproduct Data}

In the first three subsections
we list the Stasheff
local contributions to the $B_2 \wedge B_2$
cobracket component of the
two-loop heptagon NMHV ratio function, in terms of the
quantities $f_{12}$, $f_{13}$, and $f_{14}$
appearing in~(\ref{eq:nmhv7}).
Specifically, these
contain all terms of the form
$\{x\}_2 \wedge \{y\}_2$ for pairs $x, y$ having Poisson bracket
$0, \pm 1$.  The additional ``bad'' contributions to the cobracket
are shown
in~(\ref{eq:bad}) and given explicitly in the fourth subsection.

\subsection{$f_{13}$}

This function is cyclically invariant and lives entirely
in $PB_1$.
We find
\begin{align*}
\delta_{22} f_{13} &= \frac{1}{7}\bigg(\left\{\frac{\langle 1367\rangle  \langle 2347\rangle }{\langle 1237\rangle
   \langle 3467\rangle }\right\}_2\wedge \left\{\frac{\langle 1367\rangle  \langle
   2347\rangle  \langle 4567\rangle }{\langle 1467\rangle  \langle
   2367\rangle  \langle 3457\rangle }\right\}_2 \\
&-\left\{\frac{\langle 1247\rangle
    \langle 1256\rangle }{\langle 1245\rangle  \langle 1267\rangle
   }\right\}_2\wedge \left\{\frac{\langle 1245\rangle  \langle 1567\rangle
   }{\langle 1257\rangle  \langle 1456\rangle }\right\}_2 \\
&+\left\{\frac{\langle 1256\rangle  \langle 2345\rangle }{\langle 1235\rangle
   \langle 2456\rangle }\right\}_2\wedge \left(\left\{\frac{\langle 1236\rangle
   \langle 1245\rangle }{\langle 1234\rangle  \langle 1256\rangle
   }\right\}_2-\left\{\frac{\langle 1235\rangle  \langle 1567\rangle  \langle
   2456\rangle }{\langle 1257\rangle  \langle 1456\rangle  \langle
   2356\rangle }\right\}_2\right) \\
&+\left\{\frac{\langle 1247\rangle  \langle 1345\rangle }{\langle 1234\rangle
   \langle 1457\rangle }\right\}_2\wedge \left(\left\{\frac{\langle 1345\rangle
   \langle 1467\rangle }{\langle 1347\rangle  \langle 1456\rangle
   }\right\}_2-\left\{\frac{\langle 1245\rangle  \langle 1467\rangle }{\langle
   1247\rangle  \langle 1456\rangle }\right\}_2\right) \\
&+\left(\left\{\frac{\langle 1247\rangle  \langle 1345\rangle  \langle 1567\rangle
   }{\langle 1257\rangle  \langle 1347\rangle  \langle 1456\rangle
   }\right\}_2-\left\{\frac{\langle 1247\rangle  \langle 1256\rangle  \langle
   1345\rangle }{\langle 1234\rangle  \langle 1257\rangle  \langle
   1456\rangle }\right\}_2\right)\wedge \left(\left\{-\frac{\langle 1267\rangle
   \langle 1345\rangle }{\langle 1(27)(34)(
   56)\rangle }\right\}_2 \right. \\
& \qquad \left.+\left\{-\frac{\langle 1237\rangle  \langle
   1456\rangle }{\langle 1(27)(34)(56)
   \rangle }\right\}_2\right) \\
& +\left(\left\{\frac{\langle 1247\rangle  \langle 1256\rangle  \langle 1346\rangle
   }{\langle 1234\rangle  \langle 1267\rangle  \langle 1456\rangle
   }\right\}_2-\left\{\frac{\langle 1237\rangle  \langle 1345\rangle  \langle
   1567\rangle }{\langle 1257\rangle  \langle 1347\rangle  \langle
   1356\rangle }\right\}_2\right)\wedge \left\{-\frac{\langle 1234\rangle  \langle
   1567\rangle }{\langle 1(27)(34)(56)
   \rangle }\right\}_2 \bigg)  \\
& \qquad + \text{cyclic}.
\end{align*}

\subsection{$f_{12}$}

If we first define the quantity $X_1$ by
\begin{align*}
X_1 &=\left\{\frac{\langle 1367\rangle  \langle 2347\rangle }{\langle 1237\rangle
   \langle 3467\rangle }\right\}_2\wedge \left\{\frac{\langle 1267\rangle  \langle
   3467\rangle }{\langle 1467\rangle  \langle 2367\rangle
   }\right\}_2+\left\{\frac{\langle 1467\rangle  \langle 2347\rangle }{\langle
   1247\rangle  \langle 3467\rangle }\right\}_2\wedge \left\{\frac{\langle
   1347\rangle  \langle 4567\rangle }{\langle 1467\rangle  \langle
   3457\rangle }\right\}_2 \\
&-\left\{\frac{\langle 1247\rangle  \langle 1345\rangle }{\langle
   1234\rangle  \langle 1457\rangle }\right\}_2\wedge \left\{\frac{\langle
   1245\rangle  \langle 3457\rangle }{\langle 1457\rangle  \langle
   2345\rangle }\right\}_2-\left\{\frac{\langle 1457\rangle  \langle
   2347\rangle }{\langle 1247\rangle  \langle 3457\rangle }\right\}_2\wedge
   \left\{\frac{\langle 1347\rangle  \langle 4567\rangle }{\langle 1467\rangle
   \langle 3457\rangle }\right\}_2 \\
&+\left\{\frac{\langle 1256\rangle  \langle 2345\rangle }{\langle 1235\rangle
   \langle 2456\rangle }\right\}_2\wedge \left\{\frac{\langle 1236\rangle  \langle
   1245\rangle  \langle 2567\rangle }{\langle 1235\rangle  \langle
   1267\rangle  \langle 2456\rangle }\right\}_2-\left\{\frac{\langle 1267\rangle
    \langle 2356\rangle }{\langle 1236\rangle  \langle 2567\rangle
   }\right\}_2\wedge \left\{\frac{\langle 1236\rangle  \langle 2345\rangle  \langle
   2567\rangle }{\langle 1235\rangle  \langle 2367\rangle  \langle
   2456\rangle }\right\}_2 \\
&+\left(\left\{\frac{\langle 1234\rangle  \langle 1467\rangle  \langle 3457\rangle
   }{\langle 1247\rangle  \langle 1345\rangle  \langle 3467\rangle
   }\right\}_2-\left\{\frac{\langle 1245\rangle  \langle 1467\rangle  \langle
   3457\rangle }{\langle 1247\rangle  \langle 1345\rangle  \langle
   4567\rangle }\right\}_2\right)\wedge \left\{-\frac{\langle 1467\rangle  \langle
   2345\rangle }{\langle 4(12)(35)(67)
   \rangle }\right\}_2 \\
&+\left\{\frac{\langle 1467\rangle  \langle 2367\rangle  \langle 2457\rangle
   }{\langle 1267\rangle  \langle 2347\rangle  \langle 4567\rangle
   }\right\}_2\wedge \left\{-\frac{\langle 1237\rangle  \langle 4567\rangle
   }{\langle 7(16)(23)(45)\rangle
   }\right\}_2-\left\{\frac{\langle 1467\rangle  \langle 2367\rangle  \langle
   3457\rangle }{\langle 1367\rangle  \langle 2347\rangle  \langle
   4567\rangle }\right\}_2\wedge \left\{-\frac{\langle 1267\rangle  \langle
   3457\rangle }{\langle 7(16)(23)(45)
   \rangle }\right\}_2 \\
&+2 \left\{\frac{\langle 1245\rangle  \langle 2467\rangle  \langle 3457\rangle
   }{\langle 1247\rangle  \langle 2345\rangle  \langle 4567\rangle
   }\right\}_2\wedge \left\{-\frac{\langle 1234\rangle  \langle 4567\rangle
   }{\langle 4(12)(35)(67)\rangle
   }\right\}_2 \\
\end{align*}
and $X_2,\ldots,X_7$ by taking $i \to i + 1$, then we find
\begin{align*}
\delta_{22} f_{12} &= \frac{1}{7} (3,-4,3,-4,3,-4,3) \cdot (X_1, X_2, X_3, X_4, X_5, X_6, X_7) \\
&+\left\{\frac{\langle 1237\rangle  \langle 1246\rangle }{\langle 1234\rangle
   \langle 1267\rangle }\right\}_2\wedge \left(\left\{\frac{\langle 1246\rangle
   \langle 1345\rangle }{\langle 1234\rangle  \langle 1456\rangle
   }\right\}_2+\left\{\frac{\langle 1234\rangle  \langle 1467\rangle  \langle
   3456\rangle }{\langle 1246\rangle  \langle 1345\rangle  \langle
   3467\rangle }\right\}_2+\left\{\frac{\langle 1467\rangle  \langle 3456\rangle
   }{\langle 1346\rangle  \langle 4567\rangle }\right\}_2+\left\{\frac{\langle
   1246\rangle  \langle 1345\rangle  \langle 4567\rangle }{\langle
   1245\rangle  \langle 1467\rangle  \langle 3456\rangle }\right\}_2\right) \\
&+\left\{\frac{\langle 1457\rangle  \langle 3456\rangle }{\langle 1345\rangle
   \langle 4567\rangle }\right\}_2\wedge \left(\left\{\frac{\langle 1234\rangle
   \langle 1457\rangle }{\langle 1247\rangle  \langle 1345\rangle
   }\right\}_2+\left\{\frac{\langle 1234\rangle  \langle 1267\rangle  \langle
   1457\rangle }{\langle 1237\rangle  \langle 1245\rangle  \langle
   1467\rangle }\right\}_2+\left\{\frac{\langle 1237\rangle  \langle 1467\rangle
   }{\langle 1267\rangle  \langle 1347\rangle }\right\}_2+\left\{\frac{\langle
   1237\rangle  \langle 1345\rangle  \langle 1467\rangle }{\langle
   1234\rangle  \langle 1367\rangle  \langle 1457\rangle
   }\right\}_2 \right. \\
   & \qquad \left. +\left\{\frac{\langle 1257\rangle  \langle 1456\rangle }{\langle
   1245\rangle  \langle 1567\rangle }\right\}_2\right) -\left\{\frac{\langle 1256\rangle  \langle 2345\rangle }{\langle 1235\rangle
   \langle 2456\rangle }\right\}_2\wedge \left\{\frac{\langle 2567\rangle  \langle
   3456\rangle }{\langle 2356\rangle  \langle 4567\rangle }\right\}_2 \\
&+\left\{\frac{\langle 1267\rangle  \langle 2356\rangle }{\langle 1236\rangle
   \langle 2567\rangle }\right\}_2\wedge \left\{\frac{\langle 1236\rangle  \langle
   2345\rangle  \langle 3567\rangle }{\langle 1235\rangle  \langle
   2367\rangle  \langle 3456\rangle }\right\}_2 -\left\{\frac{\langle 1247\rangle  \langle 1345\rangle }{\langle 1234\rangle
   \langle 1457\rangle }\right\}_2\wedge \left\{\frac{\langle 1247\rangle  \langle
   1567\rangle  \langle 3457\rangle }{\langle 1257\rangle  \langle
   1347\rangle  \langle 4567\rangle }\right\}_2 \\
&+\left(\left\{\frac{\langle 1247\rangle  \langle 1256\rangle  \langle 1345\rangle
   }{\langle 1234\rangle  \langle 1257\rangle  \langle 1456\rangle
   }\right\}_2+\left\{\frac{\langle 1257\rangle  \langle 1347\rangle  \langle
   1456\rangle }{\langle 1247\rangle  \langle 1345\rangle  \langle
   1567\rangle }\right\}_2\right)\wedge \left\{-\frac{\langle 1247\rangle  \langle
   1567\rangle  \langle 3456\rangle }{\langle 4567\rangle  \langle 1(
   27)(34)(56)\rangle }\right\}_2 \\
&+\left(\left\{\frac{\langle 1235\rangle  \langle 2367\rangle  \langle 2456\rangle
   }{\langle 1236\rangle  \langle 2345\rangle  \langle 2567\rangle
   }\right\}_2-\left\{\frac{\langle 1235\rangle  \langle 1267\rangle  \langle
   2456\rangle }{\langle 1236\rangle  \langle 1245\rangle  \langle
   2567\rangle }\right\}_2\right)\wedge \left\{-\frac{\langle 1236\rangle  \langle
   2345\rangle  \langle 4567\rangle }{\langle 3456\rangle  \langle 2(
   13)(45)(67)\rangle }\right\}_2 \\
& \\
&+\left\{\frac{\langle 1467\rangle  \langle 3457\rangle }{\langle 1347\rangle
   \langle 4567\rangle }\right\}_2\wedge \left(\left\{\frac{\langle 1237\rangle
   \langle 1467\rangle }{\langle 1267\rangle  \langle 1347\rangle
   }\right\}_2-\left\{\frac{\langle 1267\rangle  \langle 1347\rangle  \langle
   4567\rangle }{\langle 1247\rangle  \langle 1567\rangle  \langle
   3467\rangle }\right\}_2\right) \\
&+\left\{\frac{\langle 2367\rangle  \langle 3456\rangle }{\langle 2346\rangle
   \langle 3567\rangle }\right\}_2\wedge \left(\left\{\frac{\langle 1234\rangle
   \langle 2367\rangle }{\langle 1237\rangle  \langle 2346\rangle
   }\right\}_2+\left\{\frac{\langle 1234\rangle  \langle 2367\rangle  \langle
   3456\rangle }{\langle 1236\rangle  \langle 2345\rangle  \langle
   3467\rangle }\right\}_2\right) \\
&+\frac{4}{7} \left(\left\{\frac{\langle 1257\rangle  \langle 1456\rangle  \langle
   2356\rangle }{\langle 1235\rangle  \langle 1567\rangle  \langle
   2456\rangle }\right\}_2\wedge \left\{-\frac{\langle 1235\rangle  \langle
   4567\rangle }{\langle 5(17)(23)(46)
   \rangle }\right\}_2+\left\{\frac{\langle 1357\rangle  \langle 1456\rangle
   \langle 2356\rangle }{\langle 1235\rangle  \langle 1567\rangle  \langle
   3456\rangle }\right\}_2\wedge \left\{-\frac{\langle 1567\rangle  \langle
   2345\rangle }{\langle 5(17)(23)(46)
   \rangle }\right\}_2\right) \\
&+\frac{4}{7} \left(\left\{\frac{\langle 1247\rangle  \langle 1256\rangle  \langle
   1345\rangle }{\langle 1234\rangle  \langle 1257\rangle  \langle
   1456\rangle }\right\}_2\wedge \left\{-\frac{\langle 1237\rangle  \langle
   1456\rangle }{\langle 1(27)(34)(56)
   \rangle }\right\}_2-\left\{\frac{\langle 1367\rangle  \langle 2347\rangle
   \langle 3456\rangle }{\langle 1347\rangle  \langle 2346\rangle  \langle
   3567\rangle }\right\}_2\wedge \left\{-\frac{\langle 1367\rangle  \langle
   2345\rangle }{\langle 3(17)(24)(56)
   \rangle }\right\}_2\right) \\
&-\frac{3}{7} \left(\left\{\frac{\langle 1367\rangle  \langle 1457\rangle  \langle
   2347\rangle }{\langle 1237\rangle  \langle 1467\rangle  \langle
   3457\rangle }\right\}_2\wedge \left\{-\frac{\langle 1267\rangle  \langle
   3457\rangle }{\langle 7(16)(23)(45)
   \rangle }\right\}_2+\left\{\frac{\langle 1236\rangle  \langle 2567\rangle
   \langle 3467\rangle }{\langle 1267\rangle  \langle 2346\rangle  \langle
   3567\rangle }\right\}_2\wedge \left\{-\frac{\langle 1567\rangle  \langle
   2346\rangle }{\langle 6(12)(34)(57)
   \rangle }\right\}_2\right) \\
&-\frac{3}{7} \left(\left\{\frac{\langle 1235\rangle  \langle 2367\rangle  \langle
   2456\rangle }{\langle 1236\rangle  \langle 2345\rangle  \langle
   2567\rangle }\right\}_2\wedge \left\{-\frac{\langle 1234\rangle  \langle
   2567\rangle }{\langle 2(13)(45)(67)
   \rangle }\right\}_2+\left\{\frac{\langle 1245\rangle  \langle 1467\rangle
   \langle 3457\rangle }{\langle 1247\rangle  \langle 1345\rangle  \langle
   4567\rangle }\right\}_2\wedge \left\{-\frac{\langle 1247\rangle  \langle
   3456\rangle }{\langle 4(12)(35)(67)
   \rangle }\right\}_2\right) \\
\\
&+\left(\left\{\frac{\langle 1235\rangle
   \langle 2367\rangle  \langle 4567\rangle }{\langle 2567\rangle  \langle
   3(12)(45)(67)\rangle
   }\right\}_2-\left\{-\frac{\langle 1237\rangle  \langle 2345\rangle  \langle
   4567\rangle }{\langle 3457\rangle  \langle 2(13)(
   45)(67)\rangle }\right\}_2+\frac{4}{7} \left\{\frac{\langle 1235\rangle  \langle 2367\rangle
   \langle 2457\rangle }{\langle 1237\rangle  \langle 2345\rangle  \langle
   2567\rangle }\right\}_2\right)\wedge \left\{-\frac{\langle 1267\rangle  \langle
   2345\rangle }{\langle 2(13)(45)(67)
   \rangle }\right\}_2 \\
&+\left(\left\{\frac{\langle 1237\rangle
   \langle 1345\rangle  \langle 4567\rangle }{\langle 3457\rangle  \langle
   1(23)(45)(67)\rangle
   }\right\}_2-\left\{\frac{\langle 1236\rangle  \langle 1345\rangle  \langle
   4567\rangle }{\langle 3456\rangle  \langle 1(23)(
   45)(67)\rangle }\right\}_2-\left\{\frac{\langle 1237\rangle  \langle 1456\rangle }{\langle 1(
   23)(45)(67)\rangle
   }\right\}_2\right)\wedge \left\{\frac{\langle 1267\rangle  \langle 1345\rangle
   }{\langle 1(23)(45)(67)\rangle
   }\right\}_2 \\
&+\left\{-\frac{\langle 1234\rangle  \langle 1567\rangle }{\langle 1(
   27)(34)(56)\rangle }\right\}_2\wedge
   \left(\left\{-\frac{\langle 1237\rangle  \langle 1567\rangle  \langle
   3456\rangle }{\langle 3567\rangle  \langle 1(27)(
   34)(56)\rangle }\right\}_2-\left\{\frac{\langle 1257\rangle
   \langle 1347\rangle  \langle 3456\rangle }{\langle 1345\rangle  \langle
   7(12)(34)(56)\rangle
   }\right\}_2+\frac{3}{7} \left\{\frac{\langle 1247\rangle  \langle 1256\rangle
   \langle 1346\rangle }{\langle 1234\rangle  \langle 1267\rangle  \langle
   1456\rangle }\right\}_2\right) \\
&+\left\{\frac{\langle 1237\rangle  \langle 3456\rangle }{\langle 3(12)
   (45)(67)\rangle }\right\}_2\wedge
   \left(\left\{\frac{\langle 1267\rangle  \langle 1345\rangle  \langle
   3467\rangle }{\langle 1467\rangle  \langle 3(12)(
   45)(67)\rangle }\right\}_2+\left\{\frac{\langle 1234\rangle  \langle
   1367\rangle  \langle 4567\rangle }{\langle 1467\rangle  \langle 3(
   12)(45)(67)\rangle }\right\}_2-\left\{\frac{\langle 1234\rangle
   \langle 1267\rangle  \langle 1345\rangle  \langle 4567\rangle }{\langle
   1245\rangle  \langle 1467\rangle  \langle 3(12)(45)
   (67)\rangle }\right\}_2\right) \\
\\
&+\left(\left\{\frac{\langle 1234\rangle
   \langle 1267\rangle  \langle 3457\rangle  \langle 4567\rangle }{\langle
   1247\rangle  \langle 3467\rangle  \langle 5(12)(34)
   (67)\rangle }\right\}_2+\left\{-\frac{\langle 1234\rangle  \langle 1267\rangle  \langle
   3456\rangle }{\langle 1236\rangle  \langle 4(12)(
   35)(67)\rangle }\right\}_2-\left\{\frac{\langle 4567\rangle  \langle
   3(12)(45)(67)\rangle }{\langle
   3467\rangle  \langle 5(12)(34)(67)
   \rangle }\right\}_2 \right. \\
   & \left. \qquad-\frac{3}{7} \left\{\frac{\langle 1245\rangle  \langle
   2467\rangle  \langle 3457\rangle }{\langle 1247\rangle  \langle
   2345\rangle  \langle 4567\rangle }\right\}_2-\left\{-\frac{\langle 1247\rangle  \langle 3456\rangle }{\langle 4(
   12)(35)(67)\rangle
   }\right\}_2\right)\wedge \left\{-\frac{\langle
   1234\rangle  \langle 4567\rangle }{\langle 4(12)(
   35)(67)\rangle }\right\}_2 \\
&+\left(\left\{\frac{\langle 1234\rangle  \langle 1267\rangle  \langle 3456\rangle
    \langle 3567\rangle }{\langle 1236\rangle  \langle 3467\rangle  \langle
   5(12)(34)(67)\rangle
   }\right\}_2+\left\{-\frac{\langle 1234\rangle  \langle 1267\rangle  \langle
   3567\rangle }{\langle 1237\rangle  \langle 6(12)(
   34)(57)\rangle }\right\}_2+\left\{\frac{\langle 3467\rangle
   \langle 5(12)(34)(67)\rangle }{\langle
   3456\rangle  \langle 7(12)(34)(56)
   \rangle }\right\}_2 \right. \\
   & \qquad \left.-\frac{3}{7} \left\{\frac{\langle 1246\rangle  \langle
   2567\rangle  \langle 3467\rangle }{\langle 1267\rangle  \langle
   2346\rangle  \langle 4567\rangle }\right\}_2\right)\wedge \left\{-\frac{\langle
   1267\rangle  \langle 3456\rangle }{\langle 6(12)(
   34)(57)\rangle }\right\}_2 \\
&+\left(\left\{-\frac{\langle 1267\rangle  \langle 3457\rangle }{\langle 7(
   16)(23)(45)\rangle }\right\}_2+\frac{4}{7}
   \left\{\frac{\langle 1367\rangle  \langle 1457\rangle  \langle 2357\rangle
   }{\langle 1237\rangle  \langle 1567\rangle  \langle 3457\rangle
   }\right\}_2-\left\{\frac{\langle 1467\rangle  \langle 2367\rangle  \langle
   3457\rangle }{\langle 1367\rangle  \langle 2347\rangle  \langle
   4567\rangle }\right\}_2\right)\wedge \left\{-\frac{\langle 1237\rangle  \langle
   4567\rangle }{\langle 7(16)(23)(45)
   \rangle }\right\}_2 \\
&+ \left(\left\{-\frac{\langle 1234\rangle  \langle 3567\rangle }{\langle 3(
   17)(24)(56)\rangle }\right\}_2-\frac{3}{7}
   \left\{\frac{\langle 1347\rangle  \langle 1356\rangle  \langle 2346\rangle
   }{\langle 1234\rangle  \langle 1367\rangle  \langle 3456\rangle
   }\right\}_2+\left\{\frac{\langle 1367\rangle  \langle 2347\rangle  \langle
   3456\rangle }{\langle 1347\rangle  \langle 2346\rangle  \langle
   3567\rangle }\right\}_2\right)\wedge \left\{-\frac{\langle 1237\rangle  \langle
   3456\rangle }{\langle 3(17)(24)(56)
   \rangle }\right\}_2.
\end{align*}

\subsection{$f_{14}$}

This function lives entirely in $PB_1$.  If we first define
the quantity
\begin{align*}
Y &=\left\{\frac{\langle 2347\rangle  \langle 2356\rangle }{\langle 2345\rangle
   \langle 2367\rangle }\right\}_2\wedge \left\{\frac{\langle 2346\rangle  \langle
   3567\rangle }{\langle 2367\rangle  \langle 3456\rangle
   }\right\}_2+\left\{\frac{\langle 1367\rangle  \langle 2347\rangle }{\langle
   1237\rangle  \langle 3467\rangle }\right\}_2\wedge \left\{\frac{\langle
   2347\rangle  \langle 3567\rangle }{\langle 2367\rangle  \langle
   3457\rangle }\right\}_2 \\
&+\left\{\frac{\langle 1257\rangle  \langle 1456\rangle }{\langle 1245\rangle
   \langle 1567\rangle }\right\}_2\wedge \left\{\frac{\langle 1257\rangle  \langle
   1456\rangle  \langle 2345\rangle }{\langle 1235\rangle  \langle
   1457\rangle  \langle 2456\rangle }\right\}_2-\left\{\frac{\langle 1367\rangle
    \langle 1457\rangle  \langle 2347\rangle }{\langle 1237\rangle  \langle
   1467\rangle  \langle 3457\rangle }\right\}_2\wedge \left\{\frac{\langle
   1347\rangle  \langle 4567\rangle }{\langle 1467\rangle  \langle
   3457\rangle }\right\}_2 \\
&-\left\{\frac{\langle 1237\rangle  \langle 2356\rangle }{\langle 1235\rangle
   \langle 2367\rangle }\right\}_2\wedge \left\{\frac{\langle 1236\rangle  \langle
   2567\rangle }{\langle 1267\rangle  \langle 2356\rangle
   }\right\}_2-\left\{\frac{\langle 1256\rangle  \langle 2345\rangle }{\langle
   1235\rangle  \langle 2456\rangle }\right\}_2\wedge \left\{\frac{\langle
   1235\rangle  \langle 2567\rangle }{\langle 1257\rangle  \langle
   2356\rangle }\right\}_2 \\
&+\left\{\frac{\langle 1257\rangle  \langle 1456\rangle  \langle 2345\rangle
   }{\langle 1235\rangle  \langle 1457\rangle  \langle 2456\rangle
   }\right\}_2\wedge \left\{-\frac{\langle 1235\rangle  \langle 4567\rangle
   }{\langle 5(17)(23)(46)\rangle
   }\right\}_2-\left\{\frac{\langle 1257\rangle  \langle 1456\rangle  \langle
   2356\rangle }{\langle 1235\rangle  \langle 1567\rangle  \langle
   2456\rangle }\right\}_2\wedge \left\{-\frac{\langle 1235\rangle  \langle
   4567\rangle }{\langle 5(17)(23)(46)
   \rangle }\right\}_2 \\
&+\left\{\frac{\langle 1235\rangle  \langle 2367\rangle  \langle 2457\rangle
   }{\langle 1237\rangle  \langle 2345\rangle  \langle 2567\rangle
   }\right\}_2\wedge \left\{-\frac{\langle 1267\rangle  \langle 2345\rangle
   }{\langle 2(13)(45)(67)\rangle
   }\right\}_2-\left\{\frac{\langle 1367\rangle  \langle 1457\rangle  \langle
   2347\rangle }{\langle 1237\rangle  \langle 1467\rangle  \langle
   3457\rangle }\right\}_2\wedge \left\{-\frac{\langle 1567\rangle  \langle
   2347\rangle }{\langle 7(16)(23)(45)
   \rangle }\right\}_2 \\
&+\left\{\frac{\langle 1357\rangle  \langle 2347\rangle  \langle 2356\rangle
   }{\langle 1237\rangle  \langle 2345\rangle  \langle 3567\rangle
   }\right\}_2\wedge \left\{-\frac{\langle 1237\rangle  \langle 3456\rangle
   }{\langle 3(17)(24)(56)\rangle
   }\right\}_2-\left\{\frac{\langle 1467\rangle  \langle 2367\rangle  \langle
   3457\rangle }{\langle 1367\rangle  \langle 2347\rangle  \langle
   4567\rangle }\right\}_2\wedge \left\{-\frac{\langle 1567\rangle  \langle
   2347\rangle }{\langle 7(16)(23)(45)
   \rangle }\right\}_2
\end{align*}
then we find
\begin{align*}
\delta_{22} f_{14} &=  \frac{2}{7} (Y+ \text{cyclic}) - 2 Y \\
&+\left(\left\{\frac{\langle 1257\rangle  \langle 1456\rangle }{\langle
   1245\rangle  \langle 1567\rangle }\right\}_2-\left\{\frac{\langle 1267\rangle
    \langle 2356\rangle }{\langle 1236\rangle  \langle 2567\rangle
   }\right\}_2-\left\{\frac{\langle 1257\rangle  \langle 2456\rangle }{\langle
   1245\rangle  \langle 2567\rangle }\right\}_2+\left\{\frac{\langle 1235\rangle
    \langle 1267\rangle  \langle 2456\rangle }{\langle 1236\rangle  \langle
   1245\rangle  \langle 2567\rangle }\right\}_2\right)\wedge \left\{\frac{\langle
   1256\rangle  \langle 2345\rangle }{\langle 1235\rangle  \langle
   2456\rangle }\right\}_2 \\
&+\left\{\frac{\langle 1235\rangle  \langle 2367\rangle  \langle 2456\rangle
   }{\langle 1236\rangle  \langle 2345\rangle  \langle 2567\rangle
   }\right\}_2\wedge \left(\left\{\frac{\langle 1267\rangle  \langle 2356\rangle
   }{\langle 1236\rangle  \langle 2567\rangle }\right\}_2-\left\{-\frac{\langle
   1267\rangle  \langle 2345\rangle }{\langle 2(13)(
   45)(67)\rangle }\right\}_2\right) \\
&+\left(\left\{\frac{\langle 1235\rangle  \langle 2367\rangle  \langle 2456\rangle
   }{\langle 1236\rangle  \langle 2345\rangle  \langle 2567\rangle
   }\right\}_2+\left\{\frac{\langle 1236\rangle  \langle 1245\rangle  \langle
   2567\rangle }{\langle 1235\rangle  \langle 1267\rangle  \langle
   2456\rangle }\right\}_2\right)\wedge \left\{-\frac{\langle 1237\rangle  \langle
   2456\rangle }{\langle 2(13)(45)(67)
   \rangle }\right\}_2 \\
&-\left\{\frac{\langle 1367\rangle  \langle 1457\rangle  \langle 2357\rangle
   }{\langle 1237\rangle  \langle 1567\rangle  \langle 3457\rangle
   }\right\}_2\wedge \left\{-\frac{\langle 1237\rangle  \langle 4567\rangle
   }{\langle 7(16)(23)(45)\rangle
   }\right\}_2 \\
&-\left\{\frac{\langle 1367\rangle  \langle 2347\rangle  \langle 3456\rangle
   }{\langle 1347\rangle  \langle 2346\rangle  \langle 3567\rangle
   }\right\}_2\wedge \left\{-\frac{\langle 1237\rangle  \langle 3456\rangle
   }{\langle 3(17)(24)(56)\rangle
   }\right\}_2 \\
&+\left(\left\{\frac{\langle 1367\rangle  \langle 2347\rangle  \langle 2356\rangle
   }{\langle 1237\rangle  \langle 2346\rangle  \langle 3567\rangle
   }\right\}_2+\left\{\frac{\langle 1347\rangle  \langle 2346\rangle  \langle
   3567\rangle }{\langle 1367\rangle  \langle 2347\rangle  \langle
   3456\rangle }\right\}_2\right)\wedge \left\{-\frac{\langle 1367\rangle  \langle
   2345\rangle }{\langle 3(17)(24)(56)
   \rangle }\right\}_2 \\
&-\left\{\frac{\langle 2346\rangle  \langle 3567\rangle }{\langle 2367\rangle
   \langle 3456\rangle }\right\}_2\wedge \left\{\frac{\langle 1237\rangle  \langle
   2346\rangle  \langle 3567\rangle }{\langle 1367\rangle  \langle
   2347\rangle  \langle 2356\rangle }\right\}_2 \\
&+\left(-\left\{\frac{\langle 1457\rangle  \langle 2456\rangle }{\langle
   1245\rangle  \langle 4567\rangle }\right\}_2-\left\{\frac{\langle 1567\rangle
    \langle 2456\rangle }{\langle 1256\rangle  \langle 4567\rangle
   }\right\}_2\right)\wedge \left\{\frac{\langle 1257\rangle  \langle 1456\rangle
   }{\langle 1245\rangle  \langle 1567\rangle }\right\}_2 \\
&-\left\{\frac{\langle 1457\rangle  \langle 2357\rangle  \langle 2456\rangle
   }{\langle 1257\rangle  \langle 2345\rangle  \langle 4567\rangle
   }\right\}_2\wedge \left\{-\frac{\langle 1567\rangle  \langle 2345\rangle
   }{\langle 5(17)(23)(46)\rangle
   }\right\}_2 \\
&+\left(-\left\{\frac{\langle 1567\rangle  \langle 2357\rangle  \langle
   2456\rangle }{\langle 1257\rangle  \langle 2356\rangle  \langle
   4567\rangle }\right\}_2-\left\{\frac{\langle 1567\rangle  \langle 2357\rangle
    \langle 3456\rangle }{\langle 1357\rangle  \langle 2356\rangle  \langle
   4567\rangle }\right\}_2\right)\wedge \left\{-\frac{\langle 1567\rangle  \langle
   2345\rangle }{\langle 5(17)(23)(46)
   \rangle }\right\}_2 \\
&+\left(\left\{\frac{\langle 1347\rangle  \langle 1567\rangle }{\langle
   1367\rangle  \langle 1457\rangle }\right\}_2+\left\{\frac{\langle 1567\rangle
    \langle 3467\rangle }{\langle 1367\rangle  \langle 4567\rangle
   }\right\}_2\right)\wedge \left\{\frac{\langle 1467\rangle  \langle 3457\rangle
   }{\langle 1347\rangle  \langle 4567\rangle }\right\}_2 \\
&+\left(-\left\{\frac{\langle 1347\rangle  \langle 2346\rangle  \langle
   3567\rangle }{\langle 1367\rangle  \langle 2347\rangle  \langle
   3456\rangle }\right\}_2+\left\{\frac{\langle 2346\rangle  \langle 3567\rangle
   }{\langle 2367\rangle  \langle 3456\rangle }\right\}_2+\left\{\frac{\langle
   1347\rangle  \langle 3567\rangle }{\langle 1367\rangle  \langle
   3457\rangle }\right\}_2-\left\{\frac{\langle 1347\rangle  \langle 4567\rangle
   }{\langle 1467\rangle  \langle 3457\rangle }\right\}_2\right)\wedge
   \left\{\frac{\langle 1237\rangle  \langle 3467\rangle }{\langle 1367\rangle
   \langle 2347\rangle }\right\}_2 \\
&+ \left(\left\{\frac{\langle 1237\rangle  \langle 1467\rangle  \langle 3457\rangle
   }{\langle 1367\rangle  \langle 1457\rangle  \langle 2347\rangle
   }\right\}_2+\left\{\frac{\langle 1567\rangle  \langle 2367\rangle  \langle
   2457\rangle }{\langle 1267\rangle  \langle 2357\rangle  \langle
   4567\rangle }\right\}_2+\left\{\frac{\langle 1567\rangle  \langle 2367\rangle
    \langle 3457\rangle }{\langle 1367\rangle  \langle 2357\rangle  \langle
   4567\rangle }\right\}_2 \right. \\
   & \left. \qquad +\left\{\frac{\langle 1367\rangle  \langle 2347\rangle
    \langle 4567\rangle }{\langle 1467\rangle  \langle 2367\rangle  \langle
   3457\rangle }\right\}_2\right)\wedge \left\{-\frac{\langle 1237\rangle  \langle
   4567\rangle }{\langle 7(16)(23)(45)
   \rangle }\right\}_2.
\end{align*}

\subsection{$B_1$}

Here we display the non-Stasheff local contributions to the
$B_2 \wedge B_2$ coproduct component of the two-loop
seven-point NMHV ratio function~(\ref{eq:bad}).
Exceptionally in this formula we make use of the cross-ratios
$a_{ij}$ defined in equation~(2.1) of~\cite{Drummond:2014ffa}.
We find that
\begin{align*}
B_1 &=
(a_{12} \wedge a_{16}) \wedge(a_{12}\wedge a_{61}) +
(a_{12} \wedge a_{16}) \wedge(a_{17}\wedge a_{61}) -
(a_{12} \wedge a_{23}) \wedge(a_{12}\wedge a_{61}) -
(a_{12} \wedge a_{23}) \wedge(a_{17}\wedge a_{61})
\\
&-(a_{12} \wedge a_{32}) \wedge(a_{12}\wedge a_{61}) -
(a_{12} \wedge a_{32}) \wedge(a_{17}\wedge a_{61}) -
(a_{12} \wedge a_{61}) \wedge(a_{13}\wedge a_{16}) +
(a_{12} \wedge a_{61}) \wedge(a_{13}\wedge a_{23})
\\
&+(a_{12} \wedge a_{61}) \wedge(a_{13}\wedge a_{32}) -
(a_{12} \wedge a_{61}) \wedge(a_{16}\wedge a_{23}) -
(a_{12} \wedge a_{61}) \wedge(a_{16}\wedge a_{32}) +
(a_{13} \wedge a_{16}) \wedge(a_{17}\wedge a_{61})
\\
&-(a_{13} \wedge a_{23}) \wedge(a_{17}\wedge a_{61}) -
(a_{13} \wedge a_{32}) \wedge(a_{17}\wedge a_{61}) +
(a_{16} \wedge a_{23}) \wedge(a_{17}\wedge a_{61}) +
(a_{16} \wedge a_{32}) \wedge(a_{17}\wedge a_{61})
\end{align*}
where we follow the slight abuse of notation explained
in~\cite{Golden:2014xqa}
of writing $B_1$ not explicitly as an element of $B_2 \wedge B_2$,
but rather by writing the result of the iterated coproduct acting
on $B_1$ according to $\{ a\}_2 \wedge \{ b \}_2 \mapsto
(a \wedge (1 + a)) \wedge (b \wedge (1 + b))$ and then expanding
all multiplicative terms out using the usual symbol rules.
In other words, the above formula represents the symbol
of the function $B_1$ antisymmetrized according to
$a \otimes b \otimes c \otimes d \mapsto
(a \wedge b) \wedge (c \wedge d)$.

\end{document}